\PassOptionsToPackage{table,xcdraw}{xcolor}
\documentclass[sigconf]{acmart}

\AtBeginDocument{%
  }

\copyrightyear{2024}
\acmYear{2024}
\setcopyright{acmlicensed}
\acmConference[SIGIR '24]{Proceedings of the 47th International ACM SIGIR Conference on Research and Development in Information Retrieval}{July 14--18, 2024}{Washington, DC, USA}
\acmBooktitle{Proceedings of the 47th International ACM SIGIR Conference on Research and Development in Information Retrieval (SIGIR '24), July 14--18, 2024, Washington, DC, USA}
\acmDOI{10.1145/3626772.3657747}
\acmISBN{979-8-4007-0431-4/24/07}

\usepackage{multirow}
\usepackage{amsthm}
\usepackage{enumitem}
\usepackage[normalem]{ulem}
\usepackage{color}
\usepackage{booktabs}
\usepackage{multirow}
\useunder{\uline}{\ul}{}

\newtheorem{lemma}{Lemma}

\newcommand{\ie}{\emph{i.e., }}
\newcommand{\eg}{\emph{e.g., }}
\newcommand{\etal}{\emph{et al.}}

\newcommand{\wrt}{\emph{w.r.t. }}

\begin{document}

\title{SIGformer: Sign-aware Graph Transformer for Recommendation}

\author{Sirui Chen}
\orcid{0009-0006-5652-7970}
\affiliation{
    \institution{Zhejiang University} 
    \department{The State Key Laboratory of Blockchain and Data Security} 
    \city{Hangzhou} 
    \country{China}}
\email{chenthree@zju.edu.cn}

\author{Jiawei Chen}
\orcid{0000-0002-4752-2629}
\affiliation{
    \institution{Zhejiang University} 
    \department{The State Key Laboratory of Blockchain and Data Security} 
    \city{Hangzhou} 
    \country{China}}
\email{sleepyhunt@zju.edu.cn}
\authornote{Corresponding author.}

\author{Sheng Zhou}
\orcid{0000-0003-3645-1041}
\affiliation{
    \institution{Zhejiang University} 
    \city{Hangzhou} 
    \country{China}}
\email{zhousheng_zju@zju.edu.cn}

\author{Bohao Wang}
\orcid{0009-0006-8264-3182}
\affiliation{
    \institution{Zhejiang University} 
    \department{The State Key Laboratory of Blockchain and Data Security} 
    \city{Hangzhou} 
    \country{China}}
\email{bohao.wang@zju.edu.cn}

\author{Shen Han}
\orcid{0000-0001-6714-5237}
\affiliation{
    \institution{Huazhong Agricultural University}
    \city{Wuhan}
    \country{China}}
\email{hanshen@webmail.hzau.edu.cn}

\author{Chanfei Su}
\orcid{0000-0003-3891-9672}
\affiliation{
    \institution{OPPO Co Ltd}
    \city{Shenzhen}
    \country{China}}
\email{suchanfei@oppo.com}

\author{Yuqing Yuan}
\orcid{0009-0007-6584-5433}
\affiliation{
    \institution{OPPO Co Ltd}
    \city{Shenzhen}
    \country{China}}
\email{yuanyuqing@oppo.com}

\author{Can Wang}
\orcid{0000-0002-5890-4307}
\affiliation{
    \institution{Zhejiang University} 
    \department{The State Key Laboratory of Blockchain and Data Security} 
    \city{Hangzhou} 
    \country{China}}
\email{wcan@zju.edu.cn}

\renewcommand{\shortauthors}{Sirui Chen, et al.}

\begin{abstract}
    In recommender systems, most graph-based methods focus on positive user feedback, while overlooking the valuable negative feedback. Integrating both positive and negative feedback to form a signed graph can lead to a more comprehensive understanding of user preferences. However, the existing efforts to incorporate both types of feedback are sparse and face two main limitations: 1) They process positive and negative feedback separately, which fails to holistically leverage the collaborative information within the signed graph; 2) They rely on MLPs or GNNs for information extraction from negative feedback, which may not be effective.

    To overcome these limitations, we introduce \textit{SIGformer}, a new method that employs the transformer architecture to sign-aware graph-based recommendation. SIGformer incorporates two innovative positional encodings that capture the spectral properties and path patterns of the signed graph, enabling the full exploitation of the entire graph. Our extensive experiments across five real-world datasets demonstrate the superiority of SIGformer over state-of-the-art methods. The code is available at \url{https://github.com/StupidThree/SIGformer}.

\end{abstract}





\maketitle

\section{Introduction}

Recent years have witnessed a surge of graph-based methods for recommendation \cite{ying2018graph,fan2019graph,wang2019neural,he2020lightgcn,li2023graph}. These methods generally start by creating a bipartite graph from users' historical feedback and then employ graph-enhanced representation techniques (\eg Graph Neural Network) to learn embeddings for both users and items. Owing to the graph structure's inherent capacity to encapsulate collaborative relations among users and items, graph-based methods have achieved state-of-the-art performance in collaborative recommendation.

\begin{figure}[t]
    \centering
    \setlength{\belowcaptionskip}{-5ex}
    \includegraphics[width=\linewidth]{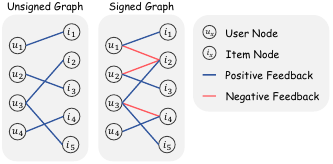}
    \caption{Illustration of sign-aware recommender system, where the signed graph can be constructed from both positive and negative feedback, carrying richer collaborative information.}
    \label{fig:sign}
\end{figure}
However, most existing graph-based methods focus on user positive feedback, while the rich negative feedback is often overlooked. In practice, negative feedback is readily available in many Recommendation Systems (RS) --- users can give low ratings, click the ``dislike'' button, or directly skip items on various platforms like Amazon, Taobao, and TikTok. This crucial negative feedback not only directly indicates users' preferences but also provides valuable collaborative information benefiting recommendation. To illustrate this point, consider Figure \ref{fig:sign}, where positive and negative feedback among users and items are constructed as a signed graph. The high-order connectivity through negative feedback also conveys useful collaborative insights. For instance, path $<u_1\mathop  {\text{---}} \limits^ -  i_2 \mathop  {\text{---}} \limits^ - u_2>$ suggests users $u_1$ and $u_2$ may have similar preferences as they both give negative feedback to item $i_2$. Additionally, the interplay of positive and negative relations offers richer collaborative relations. For example, path $<u_3\mathop  {\text{---}} \limits^ -i_4\mathop  {\text{---}} \limits^ +u_4>$ highlights the different preferences between $u_3$ and $u_4$; a longer path $<u_1\mathop  {\text{---}} \limits^ -i_2\mathop  {\text{---}} \limits^ -u_2\mathop  {\text{---}} \limits^ +i_3>$ implies that user $u_1$ is likely to favor $i_3$ as his similar user $u_2$ has previously interacted with it. 

Acknowledging the valuable collaborative information supplied by negative feedback, the integration of both positive and negative feedback presents a promising direction for enhancing graph-based recommendation. However, to the best of our knowledge, only a few studies have investigated this domain \cite{seo2022siren,liu2023pane,huang2023negative}. These methods generally construct two separate graphs from positive and negative feedback and then learn distinct representations from each, subsequently merging these representations for predictions. Despite decent performance, we identify two significant limitations:

\begin{itemize}[leftmargin=*,topsep=0pt,parsep=0pt]
    \item \textbf{The positive and negative feedback is processed separately, without a holistic consideration.} As previously discussed, the integration of positive and negative feedback within a graph offers rich collaborative information, reflecting the levels of similarity between users and items. 
        Fully exploiting such information warrants the direct utilization of the entire signed graph, rather than processing separate subgraphs independently.
    \item \textbf{The effectiveness of MLPs or GNNs in extracting information from the negative graph is questionable.}  
        Most GNNs, particularly those tailored for recommendation (\eg LightGCN \cite{he2020lightgcn}), are based on the homophily assumption --- \ie connected nodes are likely to be similar. This assumption does not hold for the negative graph. Meanwhile, MLPs struggle to fully utilize the graph structure and are challenging to train effectively in recommendation scenarios due to data sparsity.
\end{itemize}

Given the shortcomings of existing methods, we argue for the necessity of a new architecture that can fully exploit the entire signed graph. Inspired by the success of transformer architecture in many fields including language processing \cite{vaswani2017attention,kenton2019bert,brown2020language}, computer visions \cite{chen2021pre,carion2020end,dosovitskiy2020image} and sequential recommendation \cite{sun2019bert4rec,wu2020sse,fan2021continuous}, we propose leveraging transformer in this scenario. Indeed, transformer is highly aligned with the fundamental principles of collaborative filtering --- \ie estimating similarity between users and items according to their historical feedback, and then aggregating information from those similar entities for predictions.
While appealing, adapting transformer to sign-aware graph-based recommendation is non-trivial. The vanilla transformer focuses solely on semantic similarity via self-attention, lacking an explicit encoding of the collaborative information in the signed graph. Although existing graph transformer models \cite{ying2021transformers,chen2022structure,kreuzer2021rethinking} introduce subtle positional encodings to capture graph structures, they are neither specifically designed for the signed graph nor the recommendation task. To address these challenges, we introduce two novel positional encodings tailored for sign-aware graph-based recommendation:

\textbf{(1) Sign-aware Spectral Encoding (SSE).}  To integrate the structure of the entire signed graph, we propose to utilize the node spectral representation on the signed graph. Specifically, we incorporate the low-frequency eigenvectors of the signed graph's Laplacian matrix as positional encoding. Our theoretical analysis supports the efficacy of this approach: the transformer equipped with SSE can be interpreted as a low-pass filter, bringing the embeddings of user-item pairs with positive feedback closer and distancing those with negative feedback. 

\textbf{(2) Sign-aware Path Encoding (SPE).} To further capture collaborative relations among users and items, we focus on the patterns of paths within the signed graph. We encode the distance and the signs of edges along these paths into learnable parameters to capture the affinity between nodes connected by these paths. This design is based on our intuition that different path types reflect varying levels of similarity.

Equipped with these encodings, we introduce a novel recommendation method named \textbf{SI}gn-aware \textbf{G}raph Trans\textbf{former} (\textbf{SIGformer}), which adeptly utilizes the collaborative information within the signed graph. 
Its effectiveness is validated through empirical experiments on five real-world datasets, where it significantly outperforms existing graph-based methods.
Additional ablation studies further confirm the critical role of incorporating negative feedback and the efficacy of our specifically designed encodings.

Our contributions are summarized as follows:
\begin{itemize}[leftmargin=*,topsep=0pt,parsep=0pt]
    \item We highlight the importance of integrating negative feedback in graph-based recommendation and advocate for the application of transformer architecture for sign-aware graph-based recommendation.
    \item We propose two innovative sign-aware positional encodings, derived from the perspectives of signed graph spectrum and paths, which fully exploit the sign-aware collaborative information.
    \item We propose SIGformer and conduct extensive experiments to validate the superiority of our SIGformer over state-of-the-art methods. 
\end{itemize}

\section{Preliminary}
\begin{figure*}[t]
    \centering
    \includegraphics[width=0.90\linewidth]{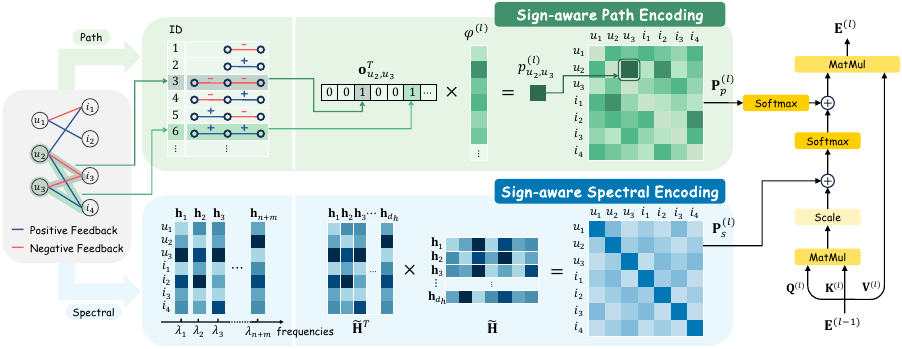}
    \caption{The illustration of our proposed sign-aware path encoding and sign-aware spectral encoding in SIGformer.}
    \label{fig:model}
    \vspace{-0.5ex}
\end{figure*}
In this section, we present the background of sign-aware graph-based recommendation and transformer model.

\vspace{-1ex}
\subsection{Sign-aware Graph-based Recommendation}

Suppose we have a recommender system (RS) with a user set $\mathcal{U}$ and an item set $\mathcal{I}$. Let $n$ and $m$ be the number of users and items in RS. User-item historical interactions can be represented as a set $\mathcal D=\{(u,i,y_{ui})|u\in \mathcal U, i \in \mathcal I\}$, where $y_{ui}=1$ signifies user $u$ has provided positive feedback on item $i$, $y_{ui}=0$ signifies negative feedback, and $y_{ui}=$`?' signifies an absence of interaction with the item. A signed bipartite graph $\mathcal{G}=(\mathcal V, \mathcal{E}^+,\mathcal{E}^-)$ is constructed from $\mathcal D$, where the node set $\mathcal V=\mathcal{U}\cup \mathcal{I}$ invovles all users and items.  The edge sets $\mathcal{E}^+$ and $\mathcal{E}^-$ correspond to user-item positive and negative interactions, respectively, \ie $\mathcal{E}^+=\{(u,i)|u\in \mathcal{U}, i\in \mathcal{I}, y_{ui}=1\}$ and $\mathcal{E}^-=\{(u,i)|u\in \mathcal{U}, i\in \mathcal{I}, y_{ui}=0\}$. The goal of sign-aware graph-based RS is to learn high-quality embeddings from the signed graph $\mathcal G$ and accordingly make accurate recommendation.

For clarity, we introduce some useful notations \wrt the signed graph. The signed graph can be partitioned into a positive graph $\mathcal G^+={(\mathcal V, \mathcal{E}^+)}$ and a negative graph $\mathcal G^-={(\mathcal V, \mathcal{E}^-)}$. Let $\mathbf A^+$ denote the adjacent matrix of positive graph $\mathcal G^+$, where each entry $\mathbf A^+_{vw}=1$ if $(v,w)\in \mathcal E^+$ or $(w,v) \in \mathcal E^+$; and $\mathbf L^+$ denote the Laplacian matrix of $\mathcal G^+$, defined as $\mathbf L^+=\mathbf I-\mathbf {(D^+)}^{-\frac{1}{2}}\mathbf A^+{(\mathbf D^+)}^{-\frac{1}{2}}$, with $\mathbf D^+$ representing the diagonal node degree matrix of $\mathcal G^+$. Let $d_u^+$ (or $d_i^+$) denote the degree of user $u$ (or item $i$) in the positive graph. Analogous definitions of notations $\mathbf A^-$, $\mathbf L^-$, $\mathbf{D}^{-}$, $\mathbf d_u^-$, $\mathbf d_i^-$ are applicable to the negative graph $\mathcal G^-$.

Compared with the traditional graph-based recommendation that only utilizes $\mathcal G^+$, sign-aware recommendation leverages the complete signed graph $\mathcal G$, encompassing both positive and negative relations. The signed graph carries richer collaborative information, necessitating effective exploitation by recommendation methods. 

\vspace{-1ex}
\subsection{Transformer}
The transformer architecture has been widely applied in many fields \cite{brown2020language,vaswani2017attention,kenton2019bert,carion2020end,dosovitskiy2020image,sun2019bert4rec}. It is composed of self-attention modules and feed-forward neural networks. In the self-attention module, the input features $\mathbf X\in \mathbb R^{n \times d}$ are projected to the corresponding query $\mathbf Q$, key $\mathbf K$, and value $\mathbf V$, and then calculated via attention with:
\begin{equation}
    \begin{aligned}
  & \mathbf Q=\mathbf X \mathbf {W}_Q, \quad \mathbf K=\mathbf X \mathbf {W}_K, \quad \mathbf V=\mathbf X \mathbf {W}_V, \\
  & \text{Attn}(\mathbf X)=\text{softmax}(\frac{\mathbf Q\mathbf K^T}{\sqrt{d_K}})\mathbf V 
    \end{aligned}
\end{equation}
where $\mathbf W_{Q} \in \mathbb{R}^{d \times d_K}, \mathbf {W}_K \in \mathbb{R}^{d \times d_K}, \mathbf{W}_V \in \mathbb{R}^{d \times d_V}$ denote the projected matrices of query, key and value respectively. 

\textbf{Connecting with Collaborative Filtering.} The transformer architecture aligns closely with the fundamental principle of collaborative filtering. Specifically, consider the input as features of users and items. The transformer initially estimates the similarity between users and items based on their projected features, then aggregates information from other entities according to this similarity, with more significant contributions from similar entities. This alignment inspires the application of the transformer in sign-aware graph-based recommendation. Nevertheless, the vanilla transformer model cannot be directly adopted, as it fails to harness the structure information of the signed graph.

\textbf{Graph Positional Encodings.} Positional Encoding has been validated as an effective solution to integrating structural information of the graph into transformer. In recent years, diverse strategies have emerged, including node degrees \cite{ying2021transformers}, shortest path distances \cite{ying2021transformers, li2020distance}, subgraph representations \cite{chen2022structure} and spectral features \cite{dwivedi2021generalization,kreuzer2021rethinking}. However, these strategies are not specifically designed for the signed graph, which requires consideration of the sign of edges. Therefore, it is imperative to develop novel positional encodings tailored for the sign-aware recommendation task.

\section{METHODOLOGY}

In this section, we first provide an overview of SIGformer (Sec 3.1), followed by a description of the proposed positional encodings (Sec 3.2 \& Sec 3.3). Finally, we elaborate on implementation details (Sec 3.4).

\vspace{-1ex}
\subsection{Overview of SIGformer}

SIGformer employs a transformer architecture for sign-aware recommendation, deviating from the conventional graph-based recommendation paradigm by replacing GNNs with transformer. Specifically, SIGformer comprises the following components:

\textbf{Embedding Module.}  As an initial step, each user and item is endowed with a $d$-dimensional embedding (\ie $\mathbf e^{(0)}_u, \mathbf e^{(0)}_i$), which can be treated as learnable parameters or transformed from user/item attributes. For a better description, we collect initial embeddings of all users and items by a matrix:
\begin{equation}
    \begin{aligned}
        {\mathbf{E}^{(0)}} = [\underbrace{\vphantom{\mathbf{e}^{(0)}_{i_1}}{\mathbf e^{(0)}_{u_1}}, \cdots ,{\mathbf e^{(0)}_{u_n}}}_{\text{user embeddings}},\underbrace{{\mathbf e^{(0)}_{i_1}}, \cdots ,{\mathbf e^{(0)}_{i_m}}}_{\text{item embeddings}}]^T.
    \end{aligned}
\end{equation}

\textbf{Sign-aware Transformer Module.} Diverging from traditional GNN-based methods, we employ a stack of multi-layer transformer to capture collaborative information. For the $l$-th layer of the transformer, the embeddings are updated iteratively as follows:
\begin{equation}
    \begin{aligned}
        {\mathbf Q^{(l)}} &= {\mathbf K^{(l)}} = {\mathbf V^{(l)}} = {\mathbf E^{(l - 1)}} \\
        {\mathbf E^{(l)}} &= \frac{1}{2}(\text{softmax} (\frac{{{\mathbf Q^{(l)}}{{({\mathbf K^{(l)}})}^T}}}{{\sqrt d }} + {\mathbf {P}^{(l)}_s}) + \text{softmax} ({\mathbf {P}^{(l)}_p})){\mathbf V^{(l)}}
    \end{aligned}
\end{equation}
Contrary to the vanilla transformer model, we omit the projected matrices  $\mathbf W_{Q}, \mathbf {W}_K, \mathbf {W}_V$ as they were found to minimally enhance performance while increasing training difficulty. Besides, we introduce two positional encodings, $\mathbf {P}^{(l)}_s$ and $\mathbf {P}^{(l)}_p$, to explicitly encode the signed graph information, which would be detailed in the next two subsections. Here we separate the two positional encodings into different softmax functions to mitigate the impact of their magnitude disparities. 

\textbf{Prediction Module.} Consistent with existing graph-based methods \cite{he2020lightgcn,yu2023xsimgcl}, After $L$ layers of transformer, we aggregate the embeddings from each layer to generate the final embeddings:
\begin{equation}
    \begin{aligned}
        \mathbf E = \frac{1}{L+1}\sum\limits_{0 \le l \le L} {{\mathbf E^{(l)}}}
    \end{aligned}
\end{equation}
The model prediction is generated from the final embeddings, \eg through an inner product, a function widely adopted by existing approaches \cite{cai2022lightgcl,seo2022siren,liu2023pane,li2023graph}:
\begin{equation}
    \begin{aligned}
        {{\hat y}_{ui}} = \mathbf e_u^T{\mathbf e_i}
    \end{aligned}
\end{equation}

\vspace{-1ex}
\subsection{Sign-aware Spectral Encoding (SSE)}

Graph spectral theory \cite{chung1997spectral,shuman2013emerging} suggests the effectiveness of spectral features (\eg Laplacian eigenvectors) in capturing the graph structure. Spectral features have also been employed to enhance the GNNs or transformer models on the vanilla graph \cite{kreuzer2021rethinking,park2022grpe,wang2022powerful}. Motivated by these successes, we propose to leverage spectral features to enhance our sign-aware transformer model.  We begin by combining the Laplacians of the positive and negative graphs as follows:
\begin{equation}
    \begin{aligned}
        \mathbf L = \frac{1}{1-\alpha}(\mathbf L^+ - \alpha \mathbf L^-)
    \end{aligned}
\end{equation}
where $\alpha$ is a flexible hyperparameter controlling the influence of the negative graph, which will be explored later. The Laplacian eigenvectors of the signed graph are:
\begin{equation}
    \begin{aligned}
        \mathbf L =  {\mathbf H^T} \boldsymbol{\Lambda} \mathbf H, \quad \mathbf {H}=[\mathbf h_1,\mathbf h_2, \cdots, \mathbf h_{n+m}]^T
        \label{eq:Laplacian eigenvectors}
    \end{aligned}
\end{equation}
where $\mathbf H$, $\boldsymbol{\Lambda}$ correspond to the eigenvectors and eignvalues respectively. The eigenvectors of the $d_h$ smallest eigenvalues denoted $\tilde {\mathbf H}$, are used for encoding node relations in the signed graph:
\begin{equation}
    \begin{aligned}
        {\mathbf {P}^{(l)}_s}= \theta^{(l)} {\tilde {\mathbf H}^T} \tilde {\mathbf H}, \quad \tilde {\mathbf H}=[\mathbf h_1,\mathbf h_2, \cdots, \mathbf h_{d_h}]^T
    \end{aligned}
\end{equation}
where $\theta^{(l)}$ is a learnable positive parameter for rescaling the magnitude.

\textbf{Connecting with Low-pass Filtering.} To elucidate the rationale behind the proposed spectral encoding, we draw a connection to low-pass filtering.  For convenience, we omit the softmax function from the analysis as its role is normalization. Similarly, for ease of discussion, we select an arbitrary column of ${\mathbf {V}^{(l)}}$ for analysis and denote it as $\mathbf v$. The vector $\mathbf v$ can be expressed by a combination of the basis $\mathbf H$:
\begin{equation}
    \begin{aligned}
        \mathbf v = \sum\limits_{1 \le k \le n + m} {{\varepsilon _k}{\mathbf h_k}}
    \end{aligned}
\end{equation}
where $\varepsilon _k$ represents the strength of the signal on the component $\mathbf h_k$. The effect of introducing ${\mathbf {P}^{(l)}_s}$ can be described as:
\begin{equation}
    \begin{aligned}
        {\mathbf {P}^{(l)}_s}\mathbf v=(\sum\limits_{1 \le k \le {d_h}} {\theta^{(l)} {\mathbf h_k}{\mathbf h_k}^T} )(\sum\limits_{1 \le k \le n + m} {{\varepsilon _k}{\mathbf h_k}} ) = \theta^{(l)} \sum\limits_{1 \le k \le {d_h}} {{\varepsilon _k}{\mathbf h_k}}
    \end{aligned}
\end{equation}
where only the low-frequency components ($h_1,h_2, \cdots, h_{d_k}$) are preserved, and higher-frequency components are filtered out. The lemma below elucidates the efficacy of this low-pass filtering nature:

\begin{lemma} The low-frequency components ($h_1,h_2, \cdots, h_{d_h}$) optimizes the following objective function:
    \begin{equation}
        \begin{aligned}
            [{\mathbf h_1},{\mathbf h_2},...,{\mathbf h_{{d_h}}}] = \mathop {\arg \min }\limits_{{\mathbf z_1},{\mathbf z_2},...,{\mathbf z_{{d_h}}}} \sum\limits_{1 \le k \le {d_h}} \Big ( & \underbrace{\sum\limits_{(u,i) \in {\mathcal E^ + }} {{{(\frac{{{\mathbf z_{ku}}}}{{\sqrt {\vphantom{d_i^+}d_u^ + } }} - \frac{{{\mathbf z_{ki}}}}{{\sqrt {\vphantom{d_i^+}d_i^ + } }})}^2}}}_{\text{Drawing Positive Neighbors}}  \\
            -\alpha & \underbrace{\sum\limits_{(u,i) \in {\mathcal E^ - }} {{{(\frac{{{\mathbf z_{ku}}}}{{\sqrt {\vphantom{d_i^+}d_u^ - } }} - \frac{{{\mathbf z_{ki}}}}{{\sqrt {\vphantom{d_i^+}d_i^ - } }})}^2}\Big ) }}_{\text{Distancing Negative Neighbors}} \\
            s.t.\quad {\mathbf z_k} \in {\mathbb R^{n + m}},\mathbf z_k^T{\mathbf z_k} = 1,\mathbf z_k^T{\mathbf z_l} = 0 , & \forall k \ne l,1 \le k,l \le {d_h} \label{eq:lemma1}
        \end{aligned}
    \end{equation}
\end{lemma}
The proof is presented in Appendix \ref{sec:proof lemma1}. From the lemma, when $\alpha>0$, the low-frequency components can be interpreted as the optimal components that minimize the distances between nodes with positive edges while maximizing the distances between nodes with negative edges. Therefore, the introduction of  ${\mathbf {P}^{(l)}_s}$ preserves those desired signals in the embeddings and filters out others, drawing the embeddings of nodes with positive edges closer together and distancing those with negative edges. The structure of the signed graph is thus explicitly encoded into the embeddings.

\textbf{The Role of $\alpha$.} This lemma also sheds light on the role of the parameter $\alpha$: it modulates the impact of the negative graph. A larger $\alpha$  implies a stronger emphasis on distancing the neighbors in the negative graph. To enhance the model's flexibility, we propose extending the range of  $\alpha$ to include negative values. Interestingly, this simple adjustment proves highly effective. 
It is based on the intuition that negative feedback may not always be really negative but instead relatively less positive compared to positive feedback \cite{huang2023negative}. For instance, in a rating system, a user's decision to leave a rating implies engagement with the item, regardless of the rating's polarity.  It suggests that the user might prefer the rated item, albeit with a low score, over others they choose not to interact with.
Consequently, it may be prudent to set $\alpha$ within the range  $(-1,1)$. Our empirical experiments also validate the optimal of $\alpha$ may be located in small negative values. 

\vspace{-1ex}
\subsection{Sign-aware Path Encoding (SPE)}

We further exploit path information within the signed graph, which explicitly reflects the collaborative relations between users and items. Our fundamental intuition is that different path types indicate varying levels of affinity between the nodes they connect.  As illustrated in Figure \ref{fig:model}, we initially enumerate all path types based on their lengths and the signs of edges within the path, assigning each path type a unique enumerated ID. This ID corresponds to a specific path type. To constrain the potentially vast space of path types, we limit our consideration to paths not exceeding a threshold length $L_p$, as excessively long paths tend to offer limited collaborative information. Consequently, the total number of path types is $N_p=2(2^{L_p}-1)$. Thereby, for any node pair $(v,w) \in \mathcal V \times \mathcal V$ in the graph, we can represent their path relationships with an $N_p$-dimensional vector $\mathbf o_{vw}$, where the $k$-th entry of $\mathbf o_{vw}$ denotes the presence or absence of the $k$-th type of path between nodes $(v,w)$. We integrate this rich path information into the transformer architecture to capture nodes' affinity:
\begin{equation}
    \begin{aligned}
        p^{{(l)}}_{vw}= \mathbf o^T_{vw}\mathbf \varphi^{(l)}
    \end{aligned}
\end{equation}
where $\mathbf \varphi^{(l)} \in {\mathbb R^{N_p}}$ is a learnable parameter capturing node affinities as reflected by the corresponding paths. Differing from existing graph transformer methods that primarily encode the shortest-distance path, our approach considers all path relations, offering a holistic view of node relations. For convenience, we aggregate $p^{{(l)}}_{vw}$ for all node pairs into a matrix, termed as sign-aware path encoding $\mathbf P^{{(l)}}_p$.

\vspace{-1ex}
\subsection{Implementation Details}

\textbf{Sampling for Acceleration.} Given the vast number of user-item combinations in Recommender Systems (RS), traversing all node pairs to calculate attention is computationally prohibitive. To address this challenge, we employ a sampling strategy. Specifically, we utilize a random walk strategy on the signed graph to pick up nodes for aggregation and concurrently record the walked path for computing $\mathbf P^{(l)}_p$. For each node $v \in V$, we perform a non-cyclic random walk of length $L_p$ starting from each neighbor of $v$ to sample a set of nodes $\mathcal S_v$ associated with the trajectory type. This allows for the rapid updating of user/item embeddings as follows:
\begin{equation}
    \begin{aligned}
        \mathbf e_v^{(l)} = \frac{1}{2}\sum\limits_{w \in {\mathcal S_v}}  \bigg( & \text{softmax} \Big(\frac{{{{(\mathbf e_v^{(l - 1)})}^T}\mathbf e_w^{(l - 1)}}}{{\sqrt d }} + {\theta ^{(l)}}m_{vw}\Big)  \\
        + & \text{softmax} ({\varphi _{{t_{vw}}}})\bigg)\mathbf e_w^{(l - 1)}
    \end{aligned}
\end{equation}
where $m_{vw}$ is the $vw$-th entry of the matrix $\mathbf M={\tilde{\mathbf H}}^T{\tilde{\mathbf H}}$, which can be pre-computed; $t_{vw}$ denotes the path type when node $w$ is sampled via the random walker. In this way, the time complexity of our attention module is reduced to $O((n+m)d\hat N)$, where $\hat N$ is the average number of nodes sampled per node. Equipped with this sampling strategy, our SIGformer achieves high efficiency. 

\textbf{Optimization.} Referring to recent work \cite{he2020lightgcn,seo2022siren}, BPR loss is adopted for optimizing our SIGformer:
\begin{equation}
    \begin{aligned}
        {\mathcal L} =  - \sum\limits_{(u,i) \in {\mathcal E^ + }} {\ln \sigma \left( {{{\hat y}_{ui}} - {{\hat y}_{uj}}} \right)}  + \sum\limits_{(u,i) \in {\mathcal E^ - }} { \ln \sigma \left( \beta({{{\hat y}_{ui}} - {{\hat y}_{uj}}} )\right)} 
    \end{aligned}
\end{equation}
where for each positive/negative feedback $(u, i)$, we sample an item $j \in \{j\in \mathcal I|y_{uj}=$`?'$\}$ that the user has not interacted with for model optimization; $\beta$ is a hyperparameter that balances the influence from the negative feedback.

\section{Experiments}

\begin{table}[t]
    \caption{Statistics of datasets, where ``Pos/Neg'' denotes the ratio between positive and negative interactions.}
    \label{tab:datasets}
    \begin{tabular}{lcccc}
        \toprule
        Dataset      & \#Users & \#Items & \#Interactions & Pos/Neg \\
        \midrule
        Amazon-CDs   & 51,267  & 46,464  & 895,266        & 1:0.22  \\
        Amazon-Music & 3,472   & 2,498   & 49,875         & 1:0.25  \\
        Epinions     & 17,894  & 17,660  & 413,774        & 1:0.37  \\
        KuaiRec      & 1,411   & 3,327   & 253,983        & 1:5.95  \\
        KuaiRand     & 16,974  & 4,373   & 263,100        & 1:1.25  \\
        \bottomrule
    \end{tabular}
    \vspace{-3ex}
\end{table}

\begin{table*}[t]
    \caption{Performance comparison between SIGformer and baselines. The best result is bolded and the runner-up is underlined. The mark `*' suggests the improvement is statistically significant with $p<0.05$.}
    \label{tab:comparison}
    \resizebox{\textwidth}{!}{%
        \begin{tabular}{cccccccccccc}
            \toprule
                                                       &                            & \multicolumn{2}{c}{Amazon-CDs}    & \multicolumn{2}{c}{Amazon-Music}  & \multicolumn{2}{c}{Epinions}      & \multicolumn{2}{c}{KuaiRec}       & \multicolumn{2}{c}{KuaiRand}      \\
                                                       &                            & Recall          & NDCG            & Recall          & NDCG            & Recall          & NDCG            & Recall          & NDCG            & Recall          & NDCG            \\
                                                       \midrule
            \multirow{4}{*}{\begin{tabular}[c]{@{}c@{}}Unsigned \\ Graph-based RS\end{tabular}}                           & LightGCN                   & 0.1325          & 0.0781          & 0.2725          & 0.1601          & 0.0854          & 0.0510          & 0.0826          & 0.0499          & 0.1197          & 0.0588          \\
                                                                                                                          & LightGCL                   & 0.1040          & 0.0591          & {\ul 0.2921}    & 0.1648          & 0.0864          & 0.0516          & 0.0848          & 0.0520          & 0.1291          & 0.0628          \\
                                                                                                                          & XSimGCL                    & 0.1346          & 0.0796          & 0.2848          & 0.1683          & 0.0887          & 0.0558          & 0.0863          & {\ul 0.0522}    & {\ul 0.1293}    & 0.0641          \\
                                                                                                                          & GFormer                    & 0.1366          & {\ul 0.0812}    & 0.2807          & 0.1648          & \textbf{0.0978} & \textbf{0.0602} & 0.0864          & 0.0520          & 0.1083          & 0.0532          \\
                                                                                                                          \midrule
            \multirow{3}{*}{\begin{tabular}[c]{@{}c@{}}Sign-aware \\ Graph-based RS\end{tabular}}                         & SiReN                      & {\ul 0.1369}    & 0.0801          & 0.2880          & {\ul 0.1725}    & 0.0804          & 0.0492          & 0.0826          & 0.0473          & 0.1167          & 0.0571          \\
                                                                                                                          & SiGRec                     & 0.1092          & 0.0648          & 0.1591          & 0.0896          & 0.0738          & 0.0475          & 0.0497          & 0.0314          & 0.1266          & {\ul 0.0699}    \\
                                                                                                                          & PANE-GNN                   & 0.1361          & 0.0810          & 0.2691          & 0.1605          & 0.0532          & 0.0301          & 0.0806          & 0.0514          & 0.1066          & 0.0522          \\
                                                                                                                          \midrule
            \multirow{2}{*}{\begin{tabular}[c]{@{}c@{}}Signed Graph\\ Embedding Methods\end{tabular}}                          & SBGNN                      & 0.0183          & 0.0100          & 0.0641          & 0.0325          & 0.0249          & 0.0143          & 0.0797          & 0.0469          & 0.0750          & 0.0361          \\
                                                                                                                               & SLGNN                      & 0.0283          & 0.0148          & 0.1498          & 0.0788          & 0.0585          & 0.0336          & {\ul 0.0865}    & 0.0508          & 0.1082          & 0.0520          \\
                                                                                                                               \midrule
            \multicolumn{1}{l}{\multirow{2}{*}{Graph Transformer}} & SGFormer                   & 0.0492          & 0.0275          & 0.2402          & 0.1373          & 0.0588          & 0.0343          & 0.0840          & 0.0504          & 0.0883          & 0.0423          \\
            \multicolumn{1}{l}{}                                   & SignGT                     & 0.0231          & 0.0121          & 0.1283          & 0.0666          & 0.0521          & 0.0300          & 0.0861          & 0.0515          & 0.0927          & 0.0439          \\
            \midrule
            \multirow{2}{*}{Our Method}                            & \multirow{2}{*}{SIGformer} & \textbf{0.1412*} & \textbf{0.0828*} & \textbf{0.3091*} & \textbf{0.1827*} & {\ul 0.0974}    & {\ul 0.0585}    & \textbf{0.0908*} & \textbf{0.0539*} & \textbf{0.1494*} & \textbf{0.0722*} \\
                                                                   &                            & +3.09\%         & +1.96\%         & +5.81\%         & +5.87\%         & -0.41\%         & -2.77\%         & +5.05\%         & +3.32\%         & +15.61\%        & +3.33\%   \\
                                                                   \bottomrule                                         
        \end{tabular}%
    }
    \vspace{-3ex}
\end{table*}

In this section, we conduct comprehensive experiments to answer the following research questions:
\begin{itemize}[leftmargin=*]
    \item \textbf{RQ1:} How does SIGformer perform compared with existing methods?
    \item \textbf{RQ2:} What are the impacts of the important components (\eg, two positional encodings, negative interactions) on SIGformer?
    \item \textbf{RQ3:} How do the hyperparameters affect the model performance?
    \item \textbf{RQ4:} How do different path types capture node similarity?
    \item \textbf{RQ5:} How does the runtime of SIGformer compare with existing methods?
\end{itemize}

\vspace{-1ex}
\subsection{Experimental Settings}

\subsubsection{Datasets}

We conduct experiments on five real-world datasets, which include both positive and negative feedback: \textbf{Amazon-CDs} \cite{mcauley2013amateurs}, \textbf{Amazon-Music} \cite{mcauley2013amateurs}, and \textbf{Epinions} \cite{tang2012etrust} are three widely-used datasets containing users' ratings on items from the Amazon and Epinions platforms. We closely refer to recent work \cite{seo2022siren,liu2023pane,huang2023negative} and consider the interactions with high ratings (\eg larger than 3.5) as positive feedback and treat others as negative. \textbf{KuaiRec} \cite{gao2022kuairec} and \textbf{KuaiRand} \cite{gao2022kuairand} record user behavior within the Kuai App. For KuaiRec, we focus on the dense dataset for experiments and classify positive and negative feedback based on the ratio of user viewing duration to total video duration. Specifically, ratios equal to or exceeding 4 are considered positive, while those below 0.1 are classified as negative. For KuaiRand, we utilize the pure version and employ ``is\_click'' attribute to classify positive and negative data as suggested by \cite{gao2022kuairand}. We adopt a conventional 5-core setting and randomly split the dataset into training set, validation set, and testing set in a ratio of 7:1:2. The dataset statistics are presented in Table \ref{tab:datasets}.

\subsubsection{Metrics}

Two widely-used metrics $Recall@K$ and $NDCG@K$ are employed for evaluating the recommendation accuracy. In this work, we simply set $K=20$ as recent work on graph-based recommendation \cite{he2020lightgcn,yu2023xsimgcl,cai2022lightgcl}.   



\subsubsection{Baselines}

To comprehensively analyze the performance of SIGformer, we compared it with various graph-based baselines:

\textbf{1) Unsigned Graph-based Recommendation Methods}. The following representative graph-based recommendation methods are included:
\begin{itemize}[leftmargin=*,topsep=0pt,parsep=0pt]
    \item \textbf{LightGCN} \cite{he2020lightgcn}: the classic graph-based method that leverages linear GNNs for recommendation. 
    \item \textbf{LightGCL} \cite{cai2022lightgcl}, \textbf{XSimGCL} \cite{yu2023xsimgcl}: the state-of-the-art graph-based methods that enhance LightGCN with contrastive learning. 
    \item \textbf{GFormer} \cite{li2023graph}: the state-of-the-art method that automates the self-supervision augmentation with transformer architecture. Considering the similarities between GFormer and SHT \cite{xia2022self}, and acknowledging that GFormer is a more recent development demonstrating better performance than SHT, we simply take GFormer for comparison.
\end{itemize}



\textbf{2) Sign-aware Graph-based Recommendation Methods.} The following methods utilize both positive and negative feedback:  
\begin{itemize}[leftmargin=*,topsep=0pt,parsep=0pt]
    \item \textbf{SiReN} \cite{seo2022siren}: the classic sign-aware recommendation method that learns two sets of embeddings from positive and negative graphs for recommendation. 
    \item \textbf{SiGRec} \cite{huang2023negative}: the representative work that analyzes the role of the negative graph and accordingly leverages GNNs to learn positive and negative embeddings. 
    \item \textbf{PANE-GNN} \cite{li2023graph}: the state-of-the-art method that leverages contrastive learning in the sign-aware graph-based recommendation model.  
\end{itemize}

\textbf{3) Signed Graph Representation Methods.} To further validate the effectiveness of our SIGformer, we include the following signed graph representation methods from the field of graph mining. We adapt these methods to recommendation tasks with additional BPR loss. 
\begin{itemize}[leftmargin=*,topsep=0pt,parsep=0pt]
    \item \textbf{SBGNN} \cite{huang2021signed}: the highly related work utilizing signed bipartite graph neural networks based on the balance principle \cite{derr2018signed}. 
    \item \textbf{SLGNN} \cite{li2023signed}: the state-of-the-art signed graph representation method leveraging signed Laplacian graph neural networks. 
\end{itemize}

\textbf{4) Unsigned Graph Representation with Transformer.} Two state-of-the-art unsigned graph transformers are included. Analogously, we adapt these methods to recommendation tasks with additional BPR loss. 
\begin{itemize}[leftmargin=*,topsep=0pt,parsep=0pt]
    \item \textbf{SGFormer} \cite{wu2023simplifying}: the state-of-the-art graph representation method leveraging a simple global attention mechanism. 
    \item \textbf{SignGT } \cite{chen2023signgt}: the state-of-the-art graph transformer method that can produce signed attention values in their attention modules. 
\end{itemize}

\subsubsection{Parameter Settings} \label{sec:parameter settings}

For our SIGformer, we adopt the Adam optimizer and search the hyperparameter with grid search. Specifically, we set the hidden embedding dimension $d$ to 64, which is alignment with recent work \cite{liu2023pane,seo2022siren}. We also draw similar conclusion with their dimensions. We simply set the learning rate to $1e-2$, the weight decay to $1e-4$, the number of eigenvectors $d_h$ to 64, and the layers of transformer to $L=3$. 
We search for $\alpha$ in the range of $[-0.8,0.8]$ with step-size 0.2, and $\beta$ in the range of $[-1,1]$ with step-size 0.2. The threshold length $L_p$ is chosen in the range of $\{1,2,3,4,5,6\}$.

For the compared methods, we use the source code provided officially and follow the instructions in the original papers to search for the optimal hyperparameters. We have traversed and frequently expanded upon, the entire hyperparameter space suggested by the authors to ensure all compared methods achieve optimal performance.

\begin{table*}[t]
    \caption{The results of the ablation study, where positional encodings or negative interactions are removed respectively.}
    \label{tab:ablation study}
    \resizebox{\textwidth}{!}{%
        \begin{tabular}{c|ccc|cccccccccc}
            \toprule
                     & \multirow{2}{*}{\begin{tabular}[c]{@{}c@{}}Negative\\ Interactions?\end{tabular}} & \multirow{2}{*}{\begin{tabular}[c]{@{}c@{}}Spectral\\ Encoding?\end{tabular}} & \multirow{2}{*}{\begin{tabular}[c]{@{}c@{}}Path\\ Encoding?\end{tabular}} & \multicolumn{2}{c}{Amazon-CDs}                        & \multicolumn{2}{c}{Amazon-Music}                      & \multicolumn{2}{c}{Epinions}                          & \multicolumn{2}{c}{KuaiRec}                           & \multicolumn{2}{c}{KuaiRand}                          \\
            \multicolumn{1}{l|}{} &                                                                                    &                                                                                &                                                                               & \multicolumn{1}{l}{Recall} & \multicolumn{1}{l}{NDCG} & \multicolumn{1}{l}{Recall} & \multicolumn{1}{l}{NDCG} & \multicolumn{1}{l}{Recall} & \multicolumn{1}{l}{NDCG} & \multicolumn{1}{l}{Recall} & \multicolumn{1}{l}{NDCG} & \multicolumn{1}{l}{Recall} & \multicolumn{1}{l}{NDCG} \\
            \midrule
            SIGformer-w/o-Neg    &                                                                                    & $\checkmark$                                                                   & $\checkmark$                                                                  & 0.1349                     & 0.0775                   & 0.2937                     & 0.1738                   & 0.0824                     & 0.0477                   & 0.0708                     & 0.0433                   & 0.1173                     & 0.0545                   \\
            SIGformer-w/o-En     & $\checkmark$                                                                       &                                                                                &                                                                               & 0.1355                     & 0.0779                   & 0.2932                     & 0.1698                   & 0.0894                     & 0.0526                   & 0.0728                     & 0.0448                   & 0.1413                     & 0.0661                   \\
            SIGformer-w/o-SPE    & $\checkmark$                                                                       & $\checkmark$                                                                   &                                                                               & 0.1380                     & 0.0798                   & 0.2988                     & 0.1744                   & 0.0959                     & 0.0574                   & 0.0862                     & 0.0520                   & 0.1471                     & 0.0697                   \\
            SIGformer-w/o-SSE    & $\checkmark$                                                                       &                                                                                & $\checkmark$                                                                  & 0.1381                     & 0.0812                   & 0.2947                     & 0.1758                   & 0.0945                     & 0.0566                   & 0.0866                     & 0.0515                   & 0.1457                     & 0.0703                   \\
            SIGformer            & $\checkmark$                                                                       & $\checkmark$                                                                   & $\checkmark$                                                                  & \textbf{0.1412}            & \textbf{0.0828}          & \textbf{0.3091}            & \textbf{0.1827}          & \textbf{0.0974}            & \textbf{0.0585}          & \textbf{0.0908}            & \textbf{0.0539}          & \textbf{0.1494}            & \textbf{0.0722}          \\
            \bottomrule
        \end{tabular}%
    }
    \vspace{-3ex}
\end{table*}

\vspace{-1ex}
\subsection{Performance Comparison (RQ1)}

The performance comparison between our SIGformer and all baselines in terms of $Recall@20$ and $NDCG@20$ is shown in Table \ref{tab:comparison}. Overall, our SIGformer outperforms all compared methods across all datasets with few exceptions. Especially in the dataset KuaiRand, SIGformer achieves impressive improvements --- 15.6\% and 3.3\% in terms of Recall@20 and NDCG@20 respectively. While SIGformer performs slightly worse than GFormer in the dataset Epinions, it still outperforms other baselines. It is worth noting that GFormer exhibits instability and even performs worse than basic LightGCN in KuaiRand.  


\textbf{Comparing with Unsigned RS.} Generally speaking, our SIGformer outperforms existing unsigned graph-based recommendation methods, while some baselines have adopted subtle contrastive learning strategies. The reason is that they overlook the negative feedback, which also provides rich collaborative information benefiting recommendation.



\textbf{Comparing with Sign-aware RS.} Our SIGformer consistently surpasses existing sign-aware graph-based recommendation methods, validating the superiority of our sign-aware transformer architecture that fully exploits the entire signed graph, as opposed to methods that separately handle positive and negative graphs. Additionally, GNNs and MLPs might not effectively extract information from the negative graph, potentially causing these sign-aware methods to underperform when compared to unsigned graph-based methods. This is evident in the performance of PANE-GNN, SiGRec, and SiReN, which are worse than LightGCN in the Epinions dataset.

\textbf{Comparing with Signed Graph Embedding Methods.}
SIGformer consistently surpasses all competing methods in signed graph representation across various datasets. Remarkably, these baselines generally exhibit subpar performance, indicating their inadequacy for recommendation tasks. This outcome can be attributed to two key factors: 1) The majority of existing signed graph methods are predicated on balance theory \cite{derr2018signed} that triads in a graph should have an even number of negative edges. Such an assumption may be overly rigid given the complex and diverse nature of user preference \cite{seo2022siren}. 2) These methods often utilize a considerable number of parameters and non-linear modules, which struggle to be effectively trained in RS due to its sparse data.


\textbf{Comparing with Graph Transformer.}
Our SIGformer still outperforms these graph transformers. This result validates the effectiveness of our SSE and SPE encodings, which are tailored for sign-aware recommendation. Existing graph transformers neither take the signed relations into consideration nor specifically designed for recommendation.  


\vspace{-1ex}
\subsection{Ablation Study (RQ2)}

\begin{figure*}[t]
    \centering
    \includegraphics[width=\linewidth]{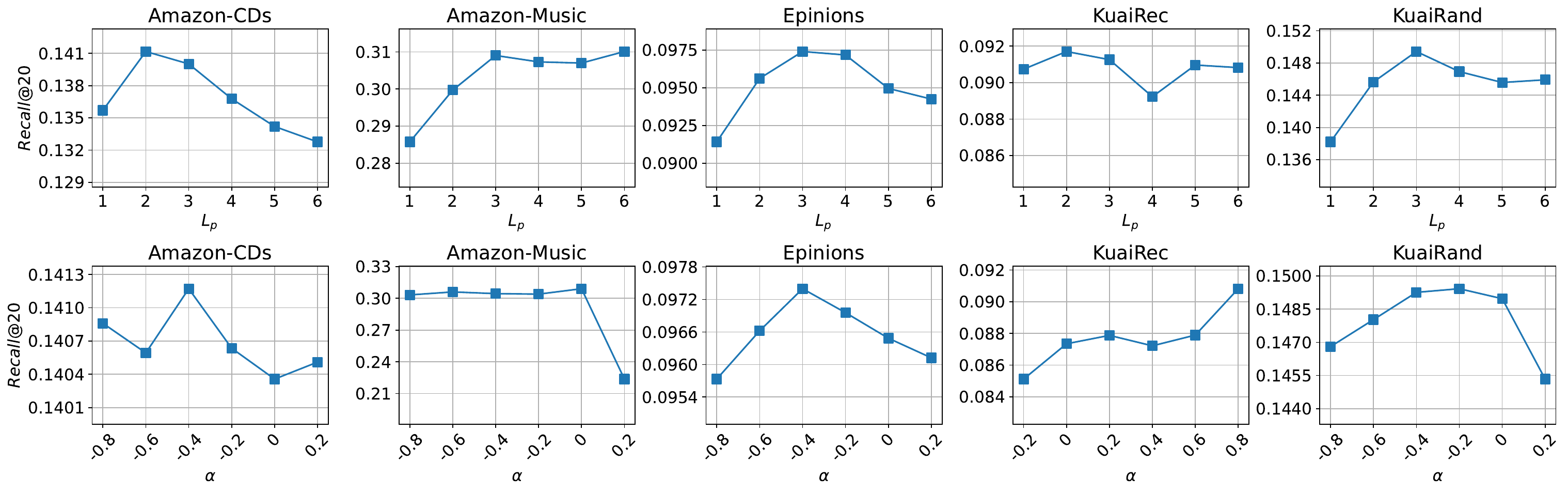}
    \caption{Performance in terms of $Recall@20$ with different $K$ and $\alpha$. }
    \label{fig:parameters}
\end{figure*}

We conduct an ablation study to investigate the effects of different modules in SIGformer. 
The results in terms of $Recall@20$ and $NDCG@20$ are shown in Table \ref{tab:ablation study}, where the two positional encodings and negative interactions are removed respectively. 

\textbf{Effects of positional encodings.} As can be seen, when removing spectral (SSE) or path (SPE) positional encodings, we consistently observe the performance drops, \ie SIGformer-w/o-SSE and SIGformer-w/o-SPE exhibit noticeably inferior performance than  SIGformer. This result clearly validates the effectiveness of our positional encodings that capture collaborative information from spectral and path perspectives.  

\textbf{Effect of negative interactions.} The removal of negative interactions from the training data results in a noticeable performance decline in our SIGformer model. This outcome reveals the significance of negative feedback, affirming our method's ability to effectively leverage its benefits.



\vspace{-1ex}
\subsection{Role of the parameters (RQ3)}

\textbf{Length Threshold $L_p$ in Sampling.} As Figure \ref{fig:parameters} shows, 
with $L_p$ increasing, the performance of SIGformer generally exhibits an initial improvement followed by a decline. This is because a larger $L_p$ provides the model with a broader receptive field, which can enhance the model's performance. However, the correlations between higher-order neighbors are weaker than those between lower-order neighbors, so an excessively large receptive field may dilute the impact of lower-order neighbors and may even bring more noise.

\textbf{Hyperparameter $\alpha$.} This hyperparameter controls the impact of negative interactions in sign-aware spectral encoding. As depicted in Figure \ref{fig:parameters}, SIGformer generally demonstrates an initial enhancement, followed by a decrement as $\alpha$ increases. This trend can be attributed to the role of $\alpha$ in balancing the impacts of positive and negative feedback --- an overly large or small influence of the negative aspect is suboptimal. Upon examining the optimal values of $\alpha$, we observe that it is located at minor negative values for the datasets Amazon-CDs, Amazon-Music, Epinions, and KuaiRand. This can be rationalized by their feedback types, namely rating and click. Although these negatively rated or unclicked items are indeed less preferred by users compared to positive ones, they might still be more favored than items with which users have not interacted. Similar conclusions have been drawn in \cite{huang2023negative}. However, in the case of KuaiRec, the negative feedback, which signifies users swiftly skipping the item, implies a strong aversion towards these items. These results validate the flexibility of our SIGformer, it can adjust the role of the negative feedback.

\vspace{-1ex}
\subsection{Case Study  (RQ4)}

\begin{figure}[t]
    \centering
    \includegraphics[width=\linewidth]{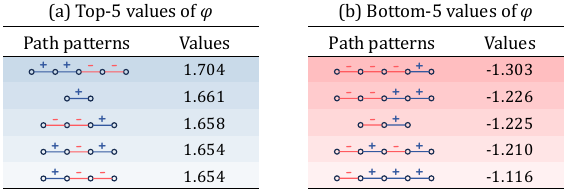}
    \caption{The top-5 and bottom-5 values of the learned $\varphi$ from KuaiRec.}
    \label{fig:SPEvalues}
\end{figure}

To investigate how SIGformer understands different path patterns, we present the top-5 and bottom-5 values of the learned $\varphi$ from KuaiRec for paths with lengths up to 4, which captures the node similarity indicated by each path type. These results are depicted in Figure \ref{fig:SPEvalues}. Here we simply choose parameters in the first layer for illustration. The majority of these results align with our expectations. For example, path types such as ``$+$'' and ``$--+$'' reflect high affinity while path types ``$---+$'' and ``$-+$'' suggest low affinity. Further, given the complexity of user preference, these results also reveal some fresh knowledge, \eg the path type ``$+-+$'' suggests strong positive relations.  

\vspace{-1ex}
\subsection{Efficiency Comparison (RQ5)}

\begin{figure}[t]
    \centering
    \includegraphics[width=\linewidth]{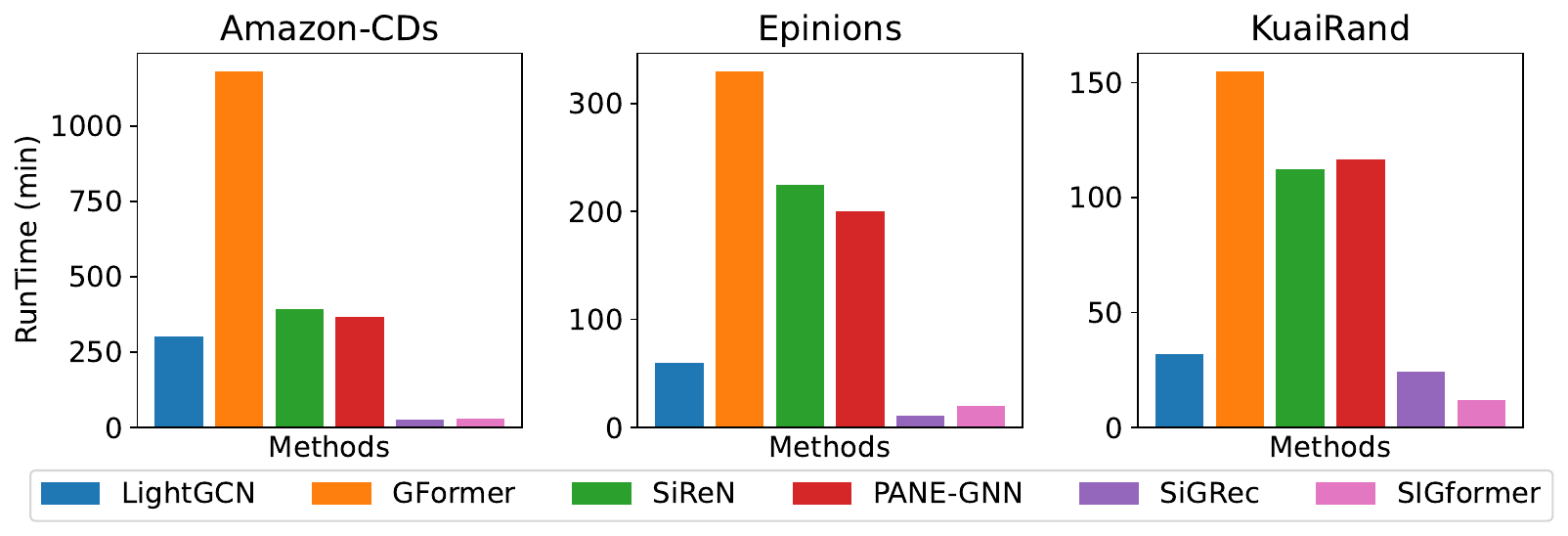}
    \caption{Runtime comparison of SIGformer with baselines.}
    \label{fig:runtime}
\end{figure}

The running time of SIGformer compared with other baselines on three large datasets are depicted in Figure \ref{fig:runtime}. We can find the transformer equipped with a random-walker-based strategy, does not impose a heavy computational burden. The efficiency of our SIGformer is comparable with SiGRec and LightGCN, and much faster than SiReN, PANE-GNN, and GFormer, which employs heavy MLPs or contrastive learning. As the random walker can be quickly implemented in GPUs, SIGformer sometimes exhibits even higher speed than LightGCN.   

\section{Related Work}
\subsection{Graph-based Recommender System}

Graph-based methods have drawn significant attention in the field of RS. Compared to traditional collaborative filtering methods such as matrix factorization \cite{koren2009matrix} and autoencoders \cite{liang2018variational}, which only utilize first-order interaction information, graph-based methods can leverage higher-order relations between nodes in user-item bipartite graph and thus exhibits better performance \cite{wu2022graph}. In the early years, Wang \etal \cite{wang2019neural}, Ying \etal \cite{ying2018graph} and Fan \etal \cite{fan2019graph} directly leveraged graph neural network in recommendation to encode the graph structure information in the representation; LightGCN further \cite{he2020lightgcn} simplified the architecture of GCN for recommendation by removing feature transformation and nonlinear activation operations, IMix\cite{deng2022graph} enhanced the generalizability of GCN by blending interacted and non-interacted item pairs for the same user; subsequently, various graph-based methods emerged that were based on contrastive learning, \eg SGL \cite{wu2021self}, LightGCL \cite{cai2022lightgcl}, SimGCL \cite{yu2022graph}, XSimGCL \cite{yu2023xsimgcl}, etc; 
other recent studies identified biases in data used for recommendations\cite{chen2023bias,chen2021autodebias} and improved existing methods from the perspective of robustness\cite{wu2023bsl,wang2024distributionally,wu2024understanding}.
The transformer architecture has also been utilized by some recent work (\eg Gformer \cite{li2023graph}, SHT \cite{xia2022self}) to improve the quality of the data augmentation. However, we highlight the following differences between our SIGformer with Gformer and SHT: 1) We directly utilize transformer as the backbone architecture, while the transformer in Gformer and SHT serves as an auxiliary role to generate augmentation; 2) our SIGformer is tailored for sign-aware recommendation, while they are designed for unsigned graph-based recommendation.      

In addition to the aforementioned graph-based methods that predominantly focus on positive data, some studies have explored leveraging both positive and negative feedback in graph-based recommendation. A pioneering example is SiReN \cite{seo2022siren}, which learned both positive and negative embeddings from corresponding graphs through GNNs and MLPs respectively, and subsequently combines these embeddings for recommendation. Huang \etal \cite{huang2023negative} conducted comprehensive analyses to reveal the role of negative feedback and developed a new method SiGRec, which enhances the learning of both positive and negative embeddings via GNNs. PANE-GNN \cite{liu2023pane} further integrates contrastive learning into negative graph representation learning. Our SIGformer advances beyond these methods in two aspects: 1) Rather than segregating the graph into positive and negative components, we harness the entire signed graph to learn embeddings; 2) We utilize transformer architecture, which is more effective in extracting collaborative information from negative feedback than MLPs and GNNs employed by these previous methods.

\vspace{-1ex}
\subsection{Sign-aware Recommender System}

The exploration of sign-aware recommendation traces its origins back to the early years of the field, with initial research predominantly focusing on \textit{explicit feedback}. {Explicit feedback}, \ie user ratings, directly indicates users' positive and negative attitudes towards items. This era saw the development of many classic recommendation methods, such as user-based CF \cite{zhao2010user}, PMF \cite{mnih2007probabilistic}, and SVD++ \cite{koren2009matrix}, etc. However, as the focus of research shifted from explicit to implicit feedback (\eg clicks, purchases), studies on sign-aware recommendation became less prevalent. Recently, however, due to the availability of negative feedback in many modern RS, sign-aware recommendation has regained significant attention. For instance, the role of negative feedback in RS has been comprehensively investigated in works such as \cite{xie2021deep,jeunen2019revisiting}. Negative feedback has also been employed to enhance various recommendation tasks, including graph-based recommendation \cite{seo2022siren,huang2023negative,liu2023pane}, negative samplers \cite{ding2019reinforced,ding2018improved}, interactive recommendation \cite{zhao2018recommendations}, and sequential recommendation \cite{pan2023understanding,park2022exploiting}, etc.

\vspace{-1ex}
\subsection{Graph Transformer}

Transformer \cite{vaswani2017attention} has been successfully applied in graph representation learning tasks. Transformer addresses key issues in GNNs, such as over-smoothing \cite{chen2020measuring,dong2021equivalence}, over-squashing \cite{alon2020bottleneck}, and limitations in expressive power \cite{xu2018powerful}, which stem from the message-passing mechanism that aggregates information from direct neighbors. However, the success of transformer in this domain usually relies on positional encodings that integrate graph structural information into the transformer framework. Recent years have seen a variety of sophisticated designs for positional encodings, including node degrees \cite{ying2021transformers}, shortest paths \cite{ying2021transformers, li2020distance}, subgraph characteristics \cite{chen2022structure}, edge relations \cite{park2022grpe}, and spectral features \cite{dwivedi2021generalization,kreuzer2021rethinking}. Other works have sought to enhance the transformer model from different perspectives. For instance, SGformer \cite{wu2023simplifying} and NAGphormer \cite{chen2022nagphormer} propose simplified graph transformer models for more efficient and effective representation learning;  while SignGT \cite{chen2023signgt} introduces signed attention values to adaptively capture diverse frequency information between node pairs. Despite these advancements, to the best of our knowledge, there remains a notable gap in transformer architectures specifically tailored for the signed graph.

\vspace{-1ex}
\subsection{Signed Graph Representation Learning}

Considering both positive and negative edges are available in many applications, signed graph representation learning draws increasing attention. Early strategies on this task include eigen-decomposition of signed Laplacian \cite{hou2003laplacian} and matrix factorization \cite{hsieh2012low}. In recent years, research has primarily relied on the balance theory that triads
in a graph should have an even number of negative edges \cite{heider1946attitudes,derr2018signed}. SGCN \cite{derr2018signed} was the first to extend GCN to the signed graph and designed a new information aggregation and propagation mechanism based on the balance theory for signed networks. SIDE \cite{kim2018side} and SIGNET \cite{islam2018signet} maintained structural balance through random walk strategies. SiGAT \cite{huang2019signed} and SNEA \cite{li2020learning} further designed graph attention mechanisms suitable for signed networks based on the balance theory. SBGNN \cite{huang2021signed} explored the balance theory in the bipartite graph and proposed a new graph neural network model for learning node representations in the signed bipartite graph. SLGNN\cite{li2023signed} returned to spectral graph theory and designed low-pass and high-pass graph convolution filters to extract low-frequency and high-frequency information on positive and negative edges. SBGCL\cite{zhang2023contrastive} introduced contrastive learning, utilizing dual-level data augmentation to capture explicit and implicit relationships among nodes in signed bipartite graphs, thereby enhancing robustness. However, these methods often can not be directly applied in RS: 1) these methods are often predicated on balance theory, which may be overly rigid given the complex and diverse nature of user preferences \cite{seo2022siren}; 2) These methods frequently employ a substantial number of parameters and non-linear modules, which struggle to be effectively trained in recommendation systems due to the inherently sparse nature of the data.

\section{Conclusion and Future Work}

This study introduces SIGformer, a novel sign-aware recommendation method that utilizes the transformer architecture to comprehensively harness the collaborative information inherent in the signed graph. Within SIGformer, we have innovatively integrated two positional encodings to capture the spectral properties and path patterns of the signed graph. Extensive experiments have been conducted to demonstrate the superior performance of SIGformer over existing graph-based recommendation methods.

A promising direction for future research is the development of a more rapid sign-aware graph transformer architecture. While SIGformer exhibits efficiency, it relies on a sampling strategy that could introduce variance and potentially affect model performance. Additionally, there is significant potential in creating more advanced positional encodings, such as those that fully leverage signed graph spectrum or incorporate additional content information, to further enhance the capabilities of the transformer in RS.

\begin{acks}
    This work is supported by the Starry Night Science Fund of Zhejiang University Shanghai Institute for Advanced Study (SN-ZJU-SIAS-001), OPPO Research Fund, the National Natural Science Foundation of China (62372399), and the advanced computing resources provided by the Supercomputing Center of Hangzhou City University.
\end{acks}

\appendix

\section{Appendices}

\subsection{The proof of Lemma 1} \label{sec:proof lemma1}
\begin{proof}
    According to the properties of symmetric normalized Laplacian,
    \begin{align}
        \mathbf z_k^T \mathbf L^+ \mathbf z_k 
& =  \mathbf z_k^T (\mathbf I-(\mathbf D^+)^{-\frac{1}{2}} \mathbf A^+ (\mathbf D^+)^{-\frac{1}{2}}) \mathbf z_k  \\
& = \mathbf z_k^T (\mathbf D^+)^{-\frac{1}{2}} (\mathbf D^+ - \mathbf A^+) (\mathbf D^+)^{-\frac{1}{2}} \mathbf z_k  \\
& = \sum\limits_{1 \leq v \leq n+m} \sum\limits_{1 \leq w \leq n+m} A^+_{vw}( \frac{\mathbf z_{kv}}{\sqrt {\vphantom{d_i^+}d_v^+}}
- \frac{\mathbf z_{kw}}{\sqrt {\vphantom{d_i^+}d_w^+}} ) ^2 \\
& = \sum\limits_{(u,i) \in {\mathcal E^ + }} 2 ( \frac{\mathbf z_{ku}}{\sqrt {\vphantom{d_i^+}d_u^+}}
- \frac{\mathbf z_{ki}}{\sqrt {\vphantom{d_i^+}d_i^+}} ) ^2
    \end{align}
    Similarly, it can be proven that 
    \begin{align}
        \mathbf z_k^T \mathbf L^- \mathbf z_k = \sum\limits_{(u,i) \in {\mathcal E^ - }} 2  ( \frac{\mathbf z_{ku}}{\sqrt {\vphantom{d_i^+}d_u^-}}- \frac{\mathbf z_{ki}}{\sqrt {\vphantom{d_i^+}d_i^-}} ) ^2
    \end{align}

    The optimization goal in Eq\eqref{eq:lemma1} can be written as:
    \begin{align}
& \mathop {\arg \min }\limits_{\mathbf z_1,\mathbf z_2,...,\mathbf z_{d_h}} 
\sum\limits_{1 \le k \le d_h} \frac{1}{2}
(\mathbf z_k^T \mathbf L^+ \mathbf z_k - \alpha \mathbf z_k^T \mathbf L^- \mathbf z_k) \\
        = & 
        \mathop {\arg \min }\limits_{\mathbf z_1,\mathbf z_2,...,\mathbf z_{d_h}} 
        \sum\limits_{1 \le k \le d_h} 
        \frac{1-\alpha}{2} \mathbf z_k^T \mathbf L \mathbf z_k \\
        = & \mathop {\arg \min }\limits_{\mathbf z_1,\mathbf z_2,...,\mathbf z_{d_h}} 
        \sum\limits_{1 \le k \le d_h} 
        \mathbf z_k^T \mathbf L \mathbf z_k
        \label{eq:optimization goal1}
    \end{align}
    where we have omitted the explicit representation of the constraint for the sake of clarity. The constraint ensures that $\mathbf z_1, \mathbf z_2,...\mathbf z_{d_h}$ form a set of $d_h$ mutually orthogonal unit vectors.
    Eq\eqref{eq:optimization goal1} represents a constrained quadratic form, which attains its minimum value when $\mathbf z_1, \mathbf z_2,...\mathbf z_{d_h}$ are chosen as the eigenvectors of $\mathbf L$ with the $d_h$ smallest eigenvalues.

    Therefore, $[\mathbf h_1,\mathbf h_2,...,\mathbf h_{d_h}]$ is a solution to Eq\eqref{eq:lemma1}.
\end{proof}

\bibliographystyle{ACM-Reference-Format}
\balance
\bibliography{main}

\end{document}